\documentclass[acmblank,letterpaper]{acmtrans2m}

\usepackage{cite}
\usepackage{array}
\usepackage{fixltx2e}
\usepackage{url}

\title{Motivation, Design, and Ubiquity: \\ A Discussion of Research Ethics \\ and Computer Science}

\author{David R. Wright \\ North Carolina State University \\ \url{drwrigh3@ncsu.edu} \\  Comments welcome --- Please send to the author at the email address above.}

\begin{abstract}
Modern society is permeated with computers, and the software that controls them can have latent, long-term, and immediate effects that reach far beyond the actual users of these systems.  This places researchers in Computer Science and Software Engineering in a critical position of influence and responsibility, more than any other field because computer systems are vital research tools for other disciplines.  This essay presents several key ethical concerns and responsibilities relating to research in computing.  The goal is to promote awareness and discussion of ethical issues among computer science researchers.  A hypothetical case study is provided, along with questions for reflection and discussion.
\end{abstract}

\date{}

\category{K. COMPUTING MILIEUX}{K.7 THE COMPUTING PROFESSION}{K.7.4 Professional Ethics}

\terms{Codes of ethics, Ethical dilemmas}

\keywords{Research ethics}

\begin{document}

\begin{bottomstuff}
This material is based upon work supported by the National Science Foundation under Grant No. 9818359. Any opinions, findings and conclusions or recommendations expressed in this material are those of the author(s) and do not necessarily reflect the views of the National Science Foundation (NSF).

Copyright 2006, David R. Wright and North Carolina State University.  Permission to make digital or hard copies of part or all of this work for personal or classroom use is granted without fee provided that copies are not made or distributed for profit or commercial advantage and that copies bear this notice and the full citation on the first page.  Abstracting with credit is permitted.
\end{bottomstuff}

\maketitle

\begin{acks}
 The author acknowledges with great appreciation Gary Comstock, Thomas Honeycutt, and Ed Gehringer for their comments and guidance in preparing this essay.
\end{acks}

\section{Introduction}\label{essay-intro}

Computers and the software that makes them work have penetrated into nearly every niche of modern life.  Manufacturing, transportation, communications, medical care, financial, and government systems, among others, are highly dependent upon computers and software systems.  This places the disciplines of Computer Science and Software Engineering in a particularly critical position, wielding more influence over the day-to-day lives of people all over the world than any other single field of research and application.  Writing about engineering in general, Bugliarello\citeyear{b:mach-mod-eth} notes that ``...ethics of engineering must go beyond broad generalities and codes of professional good conduct modeled after the Hippocratic oath.''  Little of modern engineering could be accomplished without computers and software, and computer scientists and software engineers must be aware of the broad range of ethical responsibilities that come with being direct or indirect enablers of change.

Science and engineering are commonly distinguished as two different sorts of activities.  Science, generally speaking, is the pursuit of theoretical results, while engineering seeks to apply those results through the creation and refinement of technology.  Computer science, as it is generally organized, taught, and researched, stands in a unique position, intimately involved in both theoretical and applied research.  If engineering requires an ethical position beyond professional codes of conduct, as Bugliarello notes, then a discipline that spans both theory and application, and that touches so many facets of life, should be grounded in an equally (or stronger) ethical foundation.

In discussions of ethics and computing, topics such as hacking, computer viruses, software piracy, privacy, and security often come to mind.  These subjects are commonly collected under the heading of ``Computer Ethics.''  This essay, however, is directed towards \textit{researchers} in computer science and software engineering, and is concerned with the ethical conduct of research in these and related fields.  Concern for research ethics is growing throughout the research community in general.  Given the impact research in computer science has had, and will continue to have, on a global scale, it is important that researchers in this discipline understand potential ethical dilemmas in their research.

In the LANGURE\footnote{\textit{LANGURE} is an acronym for Land Grant University Research Ethics, a NSF-funded project to develop a model curriculum in research ethics for doctoral candidates.  For more information, please see the program home page at \url{http://www.chass.ncsu.edu/langure/}.} series of modules, ethics is defined as the study of arguments about right and wrong, good and bad.  We have identified four theories that offer guidance in this decision process, and in this essay we assume that the reader has been introduced to these theories and how to apply them.  As a basis for analyzing the dilemmas we will discuss, we propose to use a simple and straightforward principle:  if everyone engaged in a particular behavior, would we consider that behavior correct and proper?  It is important to note that this perspective is concerned both with the act or behavior itself as well as the results.  Judging the rightness or wrongness of an act consequentalist and nonconsequentalist considerations is an important task in any area of research.  The reader is, of course, free to apply their own ethical analysis of the situations presented.

The intent of this paper is not only to point out concerns and potential dilemmas, but also to build awareness and to provoke discussion of these important issues within the discipline of Computer Science.  Computers have become nearly ubiquitous in modern society, and their influence reaches to the most remote parts of the globe.  Research in computer science can often find a place in real-world applications far faster than in any other discipline.  As Wulf notes, the ripple effects from innovations and technological achievements are often impossible to predict\cite{w:great-achi}, and can affect the lives of millions of people.  Furthermore, many sources for research funding require ethical review of research projects, especially when these projects involve human subjects.  In many fields the line of demarcation regarding human subjects or participants is clear.   This is often not the case in computer science research.  How this discipline responds to growing economic and regulatory pressure will directly affect researchers in computer science, and indirectly influence research in other disciplines that depend on innovations in computing for their livelihood.

The immediate goals of this essay are twofold:  first, to identify some critical ethical concerns and responsibilities relating to research in computing; and second, to promote awareness and discussion among researchers in this field about ethical concerns and issues.  Section \ref{concerns} examines practical aspects of computer science research that involve ethical judgement.  Section \ref{negl} reflects upon the preceding discussion in conjunction with the reader's own experiences and observations in an attempt to evaluate the current level of ethical awareness in the discipline, and what can (and should) be done to improve it.  A hypothetical case is presented in Section \ref{case} along with questions to stimulate discussion.  Section \ref{codes} lists Codes of Conduct for researchers and practitioners of computer science and related fields.

\section{Ethical Concerns in the Conduct of \\ Computer Science Research}\label{concerns}

In this section we discuss several key ethical issues that impact computer science research.  These are practical concerns that most researchers in the field must be aware of and address in the design, implementation, analysis, and reporting of research projects.  Most researchers in all fields are aware of the need to take special care and precautions when their research involve human subjects.  In computing research, however, there are often individuals at risk of harm that are not at all obvious.  Because of the ubiquity and casual use of computing devices today, researchers must carefully assess the possible risks to others using these devices.  Researchers must also evaluate the potential for harm to those who generate information sources subsequently used in experimental or case studies.  Identifying at-risk individuals and groups is an essential component of ethical research conduct.

The quest for knowledge and understanding is an open-ended journey, requiring researchers to look forward while building upon the work or predecessors.  Some areas of computing research are relatively mature and stable while other areas are more dynamic and fluid, and the level of stability is not necessarily dependent upon how long significant research efforts have been underway.  In these dynamic areas, researchers are often faced with a difficult choice:  corroborating prior work to strengthen the foundations of the research area or ``pushing the envelope'' while relying on prior work that may be less reliable.  The pressure to be original and cutting edge can be enormous, with the potential to make or break a career.  These decisions must be made under the influence of the basic scientific principles that define and guide the conduct of research in general.

We begin by looking at the role of software in research, both as the entity under study and as a research tool to examine and measure other entities and/or phenomena.  Potential concerns include the objectivity of the software as well as the software developer(s), interactions between different computer programs and/or computer hardware, and human interference.  These issues raise significant but often unacknowledged ethical concerns that require extra care in the design, preparation, and execution of software-related experiments.

\subsection*{Software as the subject of research}\label{soft-subj}

Software systems are often placed ``under the microscope'' of scientific study to evaluate their performance under different situations and stresses.  A research problem is identified, and a hypothesis formulated as a solution to the problem.  The hypothesis is translated into a computer program, which is then tested against the constraints that characterize the problem.  The hypothesis is either corroborated or falsified by the software tests.

To illustrate the potential ethical dilemmas involved in this research, consider the typical software development process.  A customer/client identifies a particular problem to be solved or goal to be achieved through the development of a software tool or system, for example, a web-based retail sales portal.  The problem or goal is described in detail by a set of requirements stating exactly what the software should and should not do.  In this example, the requirements would state privacy and security policies, protocols allowing the application to interact with shipping companies to calculate delivery costs, interfaces with the company's product and customer databases and accounting systems, user interface design, etc., as well as specific requirements for demonstrating that the expected improvement actually occurs.  These requirements are translated into specifications for the system, which form the basis for the design and implementation of the software.  Test suites are also developed to evaluate the correct behavior of the delivered product, based on the identified requirements.  Software that does not pass a predefined minimal set of tests is judged to be unacceptable because it does not fulfill a minimal set of requirements.

While a research hypothesis usually will not provide a full set of requirements in a formal and explicit manner, those requirements exist implicitly in the description of the problem, the context in which the problem is being investigated, and the solution proposed by the hypothesis.  An example might involve the development of a new algorithm for searching databases.  The researcher's hypothesis states that this new algorithm resolves a problem existing algorithms exhibit when searching for particular kinds of information in databases using certain types of structure or organization.  The hypothesis specifies an algorithm (with specific requirements as to its implementation), a particular kind of search operation (characterizing the contents of the databases), and the structure and organization of those databases.  These requirements define the functional and nonfunctional behaviors and characteristics of the research software just as they do in a commercial development project.  This leads to two questions:

\begin{enumerate}
 \item If the software under study appears to corroborate the hypothesis, is this due to a weak suite of tests, a weakly stated hypothesis, ``creative programming,'' or accurate corroboration?
 \item If the software under study appears to falsify or disprove the hypothesis, is this a result of an error in creating the software (or test suite), an intentional effort to disprove the hypothesis, or an accurate falsification?
\end{enumerate}

Answering these questions is not a trivial task, and will require careful examination of experimental design, software development, environmental constraints, data collection and analysis, and duplicability.  Computers are fundamentally deterministic machines:  they do what they are told to do.  If the results are not what was expected, is it because the expectation was wrong, or does the fault exist in the instructions given?  Similarly, if the end results correspond with the expectations, is it because the expectations were valid, or was a particular set of instructions chosen that would lead to those results regardless of the validity of the hypothesis?

Defects in software are the norm rather than the exception, and most of the research in software engineering over the past forty years has sought processes to develop software systems with fewer defects, and techniques to detect defects more reliably.  Complicating the defect level is the difficulty in exhaustively testing software systems to guarantee ``zero defects'' in the delivered product.  Based on the historical track record of software development, an unexpected falsification of a research hypothesis is often viewed with a degree of skepticism, spurring more careful examination of the software under study and the environment in which the failure occurred.  This may then lead to a correction of defects in the software or test suite, or a re-evaluation of the hypothesis.

Similarly, programming errors can also result in a false positive experimental result.  The software under study may not accurately reflect the requirements implied in the hypothesis,  or the test suite (experimental setup) may not accurately measure the results, either accidentally or intentionally.  Researchers must apply the same level of skepticism and caution to justify positive results as they do to mitigate a negative outcome.  Furthermore, detailed records and documentation must be maintained in order to justify the correctness of the experimental approach, software design and implementation, experimental environment (hardware and software), and collected data\cite{z:rel-res}.

In the same paper, Zobel points out several other related ethical problems in empirical studies of software, noting that the actual implementation code itself is not sufficient documentation of an experiment.  A researcher's bias towards a particular algorithm or implementation can alter the results of an experiment.  The choice of data, underlying hardware and software, and the quality of the implementations used in an experiment can have significant effects on the results.  Programmer skill and experience with the hardware environment can also have critical effects:  a programmer with minimal experience using the C programming language on a UNIX or Linux machine may write less reliable code than one with more extensive experience.  He also notes that it is not sufficient to note that a particular implementation worked for a particular data set; it must be shown to work consistently on a variety of input data.  Additionally, the experimenter must explain \textit{why} the implementation worked, as well as providing detailed information about failures\cite{z:rel-res}.  These are among the standards for research in most other scientific disciplines, and certainly should be in computer science.

Researchers must also exercise care in the selection and application of metrics and analysis methods.  The search for reliable metrics continues to be an active research area, motivated, in part, by the need to define a core set of measurements and analyses to measure various aspects of software quality and performance.  El Emam, Benlarbi, Goel, and Rai demonstrated that even commonly accepted metrics are not as reliable as expected, identifying a ``confounding effect'' in certain object-oriented source code metrics\citeyear{ebgr:valid-oo-metric}.  These metrics had been shown to correlate with the \textit{fault-proneness} of a class implementation, but using the method they developed to control for the class implementation size, the metrics no longer showed the expected relationship.  Repeating prior experiments by other researchers generated the same results (i.e., the lack of correlation between the metrics and the defects), leading them to question the common use of these metrics and to recommend revisiting previous studies to confirm earlier conclusions.

Metrics and analysis methods are primary instruments of computer scientists researching software systems.  The development and validation of these metrics and methods is complicated by the dynamic and diverse nature of computing itself:  metrics are usually dependent upon the programming language used and the platform where the software is executed.  A lack of consensus on the topic of measuring instruments and metrics combined with minimal attention directed towards the quality of measurement results in this discipline, as compared to other science and engineering fields, further complicates the researcher's work \cite{as:meas-se}.  Jones \citeyear{j:sw-metr} is even more critical:  ``The software industry is an embarrassment when it comes to measurement and metrics,'' noting that the measurements generally reported in the literature lack the precision to duplicate the author's work, a canonical requirement in other scientific disciplines.

Because software is a human-created entity, and is fundamentally deterministic (although this determinism may be obscured by the system's complexity), many opportunities for bias, both intentional and accidental, exist in the study of software itself.  Furthermore, computer software is an artifact of human creation that exists (as a dynamic phenomenon) within another artificial, human-created environment, and researchers must be careful to ensure the objectivity of experimental designs and implementations.  Software written for the expressed purpose of demonstrating a particular performance characteristic should be as neutral as possible, and where implementation biases exist, they \textit{must} be disclosed and their effect on the results explained to allow other researchers and practitioners to relate those results to their own work --- this is a hallmark of responsible science.  Well-known and understood metrics should be used whenever possible, and any possible biases in measurement, such as those identified in \cite{ebgr:valid-oo-metric}, should also be accounted for in the analysis of the results.  The failure to disclose biases, either actual or potential, in the implementation, metrics, or analysis of experimental software can easily be interpreted as laziness or even as an attempt to intentionally deceive others, neither of which are acceptable in scientific research\cite[p. 168]{nas:resp-sci-v2}.

Detailed record-keeping is an essential aspect of scientific research, but one that is often neglected in software research.  Part of this is attributable to the feeling that ``the records are in the software'' and associated documentation\cite{z:rel-res}.  As Zobel goes on to point out, this is a very weak assumption on several accounts.  The current version of the software, while a result of all the changes that occurred in prior versions, usually does not retain all of the original code.  The relationships between the source code, software requirements and design, experimental design, and research hypothesis can easily be obscured, and the source code itself cannot capture and store the entire sequence of design and implementation decisions that led to it:  the finished product is a \textit{result} of all of those decisions, but not a \textit{record} of them.  Zobel's recommendation is to adopt the practices of other research disciplines, keeping detailed daily journals of all research-related work, use version control systems to maintain all versions of experimental software, and to record detailed logs of all experimental data, even if there is no expectation of using all of the data generated\cite{z:rel-res}.  The nature of software and the virtual environment in which it exists does not permit short-cutting the essential tasks of researchers.  Rather, this nature and environment require as much, if not more care, discipline, and effort than may be common in other fields.

\subsection*{Software as a research tool}\label{soft-tool}

In addition to being the subject of investigation, software is used in computer science research as a tool for making a variety of measurements in existing or new systems, either hardware or software.  The use of software benchmarks (programs that measure the performance of hardware and software systems) became popular in the early 1980s, spurred by the variety of hardware platforms resulting from the introduction of personal computer systems. These results generated by these tools were frequently used in commercial advertising as well as in research environments.  Benchmarking software continues to be a widely used method for evaluating hardware and software system performance, particularly in research environments. The problems surrounding the choice of measuring instruments, how the measurements are actually implemented in the system under study, and the interpretation of the results raise several ethics-related issues.

We begin a discussion about the ethical use of software instrumentation with several concerns, based on the work of Levy and Clark\citeyear{lc:bench-perf}, regarding the choice of benchmark software. First and foremost, a researcher must understand what it is that the benchmark program actually measures, as well as the specific behavior(s) or attribute(s) to be measured. Pointing out weaknesses in benchmark software, Gustafson identifies several examples of benchmarks that do not necessarily measure what the user might expect. One example checks the results of floating-point calculations to only four decimal places, although the benchmark program itself requires the use of high-precision (64-bit) data\cite{g:purp-bench}. This approximation saves processing time, but if the user is interested in floating-point mathematical operations and is not aware of this benchmark's limitations, the interpretation of the results may be in error. Investigating the \textbf{\texttt{health}} benchmark, Ziles found that this particular tool was not a valid benchmark for the type of measurement it performed, but rather was a ``micro-benchmark'' and an inefficient one at that\citeyear{z:bench-harm}.  The \textbf{\texttt{health}} benchmark is a health care simulator that models a hierarchy of hospitals and the patients undergoing treatment at each one, and is used to evaluate the performance of linked data structures.  Ziles found that the benchmark actually measured traversal time over long linked-lists, and demonstrated an algorithmic modification to the benchmark itself that improved the execution time of the tool by a factor of 200.  Furthermore, he also found that the model itself is flawed because the number of patients arriving exceeds the capacity of the hospitals to treat and discharge them; thus as the simulation is run for longer times to collect more data, the reported performance of the system under evaluation will decline as the waiting list grows longer and longer.

A second area of concern that Levy and Clark identify is that of bias in benchmarks as a result of how the programs are implemented, including the particular programming language and/or compiler used to create the program\cite{lc:bench-perf}. Programming languages have different features and semantics that may offer more efficient ways of coding specific operations. Compilers are platform-specific, and may also perform optimizations tailored to specific hardware architectures. While a researcher might be able to easily recognize optimizations resulting from the choice of programming language or particular features within a language, identifying architecture-specific optimizations performed by a compiler is a much more difficult task. Levy and Clark highlight an example of this kind of bias using several different benchmark programs written in different languages, and compiled and executed on different hardware platforms\cite{lc:bench-perf}.  Their goal was to compare the performance of these tools to identify differences related to the implementation language and execution platform.  They concluded that the benchmark performance cannot be attributed to any one factor, and that this type of scalar measurement of execution time is not a reliable way to compare the performance of different hardware and software systems.  A direct comparison of two different compilers, using the same hardware architecture and platform, was done recently by Gurumani and Milenkovic, who found that code compiled by the Intel C++ compiler performed better than code generated by Microsoft's Visual C++ compiler using a recognized standard benchmark suite\cite{gm:cpp-bench}.

The Standard Performance Evaluation Corporation (SPEC) CPU2000 suite of benchmarks is an industry-standardized reference for measuring a computer system's processor, memory, and compiler support\cite{spec:cpu2000}.  Citron surveyed 173 papers published in three respected computer architecture conferences two years after the suite's introduction.  He found that 115 of these papers cited the use of of a SPEC benchmark, but only 23 of these used then entire suite\cite{c:misspec}.  Of the 92 partial uses, only 27 papers discussed the reason(s) for not using the entire suite.  This survey raises two concerns:  the limited use of the entire suite of benchmarks that are widely accepted as stable and reliable performance measuring instruments; and the lack of explanation for omitting some of the tests or substituting simulated results in place of actual measurements.  Citron notes that this partial use can be misleading to readers, particularly when little or no explanation is given --- readers may not have sufficient information to draw realistic conclusions about the research results.

Hennessy, Citron, Patterson, and Sohi\citeyear{hcps:use-abuse-spec} continue this discussion, generally drawing the same conclusions regarding the misuse of the SPEC benchmarks as Citron did in the paper noted above.  Sohi, however, argues that there are sometimes extenuating circumstances for not using the entire suite:  not all of the tools compile, certain tools evaluate aspects of performance not thought to be related to the problem at hand, etc.  He compares this to pharmaceutical researchers who do not try to evaluate a new drug's performance against all possible diseases.  This statement is misleading on two key points.  First, as Hennessy points out in the conclusion of this discussion, alternative applications are sometimes found for drugs that are not effective for their original purpose, citing Viagra as an example.  Second, Sohi does not recognize that most of the research performed on a new drug prior to [human] clinical trials is designed to determine the positive and negative effects of the drug\cite{fda:beg}.  Concern for negative side effects continues throughout the clinical testing phase\cite{fda:clin}, and throughout the life of the drug, as exhibited by the recent removal of an entire group of drugs (including Vioxx, Celebrex, and Bextra) from the market as a result of significant risks to patients\cite{fda:cox2-talk}.  Unfortunately, Sohi seems to ignore the fact that thorough testing can illuminate negative behaviors in computer systems that limited benchmarking may hide.

The third issue that computer scientists who are using benchmarks need to be aware of is how
the benchmark program interacts with the system under study. On one hand, a program that
consumes excessive resources such as memory, processor time, or disk I/O may significantly affect the measurements made. On the other hand, if the benchmark does not interact sufficiently with the system under study, it may not be able to accurately collect data from the system. The compromise usually involves some kind of data sampling coupled with statistical analysis. Often, this analysis is the computation of a mean value for a data set, although the choice of which mean (arithmetic, harmonic, or geometric) to use is sometimes a matter for dispute. Mashey concludes that the choice of which mean to use for a particular benchmark is less important than understanding the statistical distribution of the data set\cite{m:war-bench}.

Benchmarking tools are a necessary component of computer science research and can
effectively and positively influence the maturity of the discipline by providing a common set of experimental evaluations tools and techniques.  The availability of standardized, well-defined means of making measurements is an essential element of science and research. This common ground fosters consensus, collaboration, and rigor\cite{seh:swe-res-bench}. It is critical, however, that the tools themselves are well understood and documented so that researchers can choose the instrument appropriate for the task at hand. Researchers must also take responsibility for the benchmarks they choose to employ, and fully disclose the reasons for making those choices as well as detailed explanations of how the tools are used. It is also essential to accurately and specifically define \textit{what} is measured and why.

\subsection*{Basic principles of scientific inquiry}\label{basic-prin}

Rational discourse, replication, and criticism are cornerstones of modern scientific inquiry, providing the means to advance science and to translate theory into practice and technology.  They also help define what is and is not responsible conduct and reporting of research, a critical topic for a relatively young discipline such as computer science.  We will consider each of these three subjects as they apply to computer science research and discuss how an awareness of their ethical importance can improve and strengthen research efforts in this discipline.

Rational discourse is the clear and commonly understood communication among scientists of a discipline, and between them and researchers in other disciplines and the ``outside'' world.  On the surface, this communication involves style and standards for publication, and computer science has built largely upon earlier standards established in fields like mathematics and logic.  It is important, however, to look beyond the \textit{how} we communicate and examine \textit{what} we discuss and \textit{why}.  Computer science has evolved from its origins in mathematics and logic to encompass a broad diversity of research areas, ranging from formal languages and models of computation, stochastic processes, algorithms, and data structures to human-computer interaction, networked and embedded computing, and the many specialties that fall under the umbrella of software engineering.  Over the past fifty years, computers have moved from the isolated and controlled ``computer room'' to a nearly ubiquitous presence in modern life.

This variety of perspectives has led to a corresponding diversity of research in computer science.  Without a well-defined body of knowledge, these ``branches" have fed primarily on themselves in terms of prior foundational work. While diversity of opinion and research emphasis is not a sign of poor ethical judgement or practice, relying strictly on one path of research introduces additional risk into the research effort itself, particularly when references to work outside that perspective are the exception rather than the rule in new research. Precisely because the body of knowledge is still youthful and growing, it is critically important to relate new research to other investigations into the same or similar problems. It is also important to seek out and relate research in fields outside software engineering to this new work, as results in other fields may have direct and valuable bearing on computer science research.

In his 1993 Turing Award Lecture, Hartmanis argues that computer science is fundamentally different from other sciences because it does not conform to the paradigms of the physical sciences \cite{h:turing}.  The ``defining characteristics of computer science'' include the unprecedented difference in scale between individual bits of data and programs and the billions of instructions or operations carried out each second by modern computers and the different roles played by theory and experiment.  Theories are used to develop methodologies, models, logics, and semantics to be used on program design and implementation, rather than to explain observations and predict new phenomena.  Experiments in computer science demonstrate the validity of these methods and models, constrained by their design contexts and actual implementations. Hartmanis concludes that computer science is indeed an ``independent new science, but it is intertwined and permeated with engineering concerns and considerations'' and a ``new form of engineering...the engineering of mathematics or mathematical processes.''\cite{h:turing}  Because of these differences, Hartmanis contends that research in computer science, and the rules of more ``traditional'' sciences do not equally apply, in particular to empirical research.

Stewart counters Hartmanis's arguments by first noting that physics deals with multiple scales:  phenomena ranging from the tiniest subatomic particles to the entire universe\cite{s:sci-cs}, a span of at least $10^{41}$ in magnitude and eclipsing the ``immense differences in scale'' that Hartmanis claims are present in computer science\cite{h:turing}.  Stewart also notes that many research areas in computer science are equally capable of developing theories that can be empirically judged rather than just demonstrated, suggesting the common relation between theory and experiment in this discipline is a ``fatal flaw to be remedied'' rather than a reason for maintaining the \textit{status quo}\cite{s:sci-cs}.  Stewart agrees that computer science is a science in the traditional sense, while sternly criticizing the discipline for its failure to adhere to the basic principles of modern science.  In particular, he notes that many subdisciplines rely more on exploratory experimentation rather than stating well-defined problems amicable to theoretical and empirical analysis.

Freeman asserts that an effective computer science research program requires not only a strong core knowledge, but must also interact with other disciplines in a larger, real-world context \cite{f:eff-cs}.  Hartmanis also asserted that computer science must ``increase its contact and intellectual interchange with other disciplines,'' look for important emerging computer science problems in other fields, avoid focusing on distinctions between basic and applied research and development, and embrace the transfer of knowledge between academic, industrial and social researchers\cite{h:comp-fut}.   Leveson has noted the need to incorporate research in social and cognitive sciences into software engineering research\cite{l:se-lim-comp}.  Glass, Vessy, and Ramesh also concluded that software engineering research is both narrow in research method and approach, and focuses on itself as a reference discipline rather than looking outward for foundational or supportive research\cite{gvr:rsch-se}.  Given the current ubiquity of computing and the prospect that it will become more and more integrated into society, it is imperative that research topics and discourse become more open and active in interdisciplinary research efforts.  Computer science must build not only upon itself, but must actively seek out and build upon research and theory from other disciplines.  Failure to do so is a breach of the implied moral and social contract between science and society.

The ability to duplicate the work of other researchers is perhaps the most fundamental principle and responsibility of science.  Repeating an experiment allows a new result to be corroborated or refuted, as well as providing the means to restate and refine the problem under consideration.  Duplicating the prior work of other researchers is often also more than simply recreating the earlier experiment:  the later researcher should also be looking for new results that extend or clarify the earlier work or have stated a thesis outlining why the duplication is expected to fail.  It is the continual refinement of hypotheses that builds credibility of researchers and results alike.

Rational discourse is a requisite for replicating research.  Clear and precise descriptions of experimental setups and protocols, research hypotheses, results anticipated based on theoretical analyses, and complete records of collected data provide the groundwork from which other researchers can attempt to corroborate, refute, and refine existing research.  From these attempts to duplicate prior work new, and potentially more interesting, problems become clear.  Detailed records can also protect researchers against accusations of fraud or misconduct, just as lapses may indicate their presence.

Given the diversity of research topics within computer science, there cannot be a single standard of judgement for documentation and record keeping.  Some experiments, e.g., those studying human reactions or interactions with computing systems, will have requirements closer to those of psychological or cognitive science research, while areas such as experimental algorithmics will not require the same degree of detail\cite{z:rel-res}.  Zobel also points out that, in light of rapidly changing computer technology, it may be nearly impossible to experimentally reproduce results, but that these results should illustrate the same underlying phenomena.

It is the responsibility of computer science researchers to build upon the work of others, while creating new knowledge and understanding to extend and strengthen the discipline.  For a young discipline or subarea, it is often critical to look outside the immediate bounds of that research area to similar fields in other areas of investigation.  Unfortunately, that outward-looking attitude is the exception rather than the rule in computer science and software engineering:  89\% and 98\% (respectively) of literature references were within the field of study according to one survey of publications\cite{grv:rsch-comp-disc}.  Not only does this introversion place limits on possible insights and innovations, it reflects a separation from the larger community of science and society.

This leads us to the issue of criticism within the rational discourse of a discipline.  Science grows through critical analysis and testing of hypotheses:  one need only look at the history of science to see this phenomenon.  But critical analysis is only effective when hypotheses are stated clearly and precisely.  Generalizations may be easy to corroborate, but ultimately tell us little about the real world, since they can be very difficult to disprove.

Popper identifies four critical lines of testing\cite[pp. 32-33]{p:logic-sci-disc} for scientific theories:
\begin{itemize}
 \item The comparison of conclusions among themselves for internal consistency.
 \item Investigation of a theory's logical form to determine if it is empirical/scientific or tautological.
 \item Comparison with other theories to determine if it represents a scientific advance.
 \item Testing via empirical application of conclusions derived from the theory.
\end{itemize}

The identification of these types of testing lead to his definition of the term \textit{falsifiability} as the ability for a scientific system to be refuted by experience\cite[pp. 40-41]{p:logic-sci-disc}.  Basically, he asserts that scientific theories are not ``proven true'' by positive results, but only supported temporarily pending a negative result.  He also shows that the \textit{empirical content} of a statement increases with the degree of falsifiability --- the more specific a theory is (e.g., the more it forbids), the more that theory says about the world of experience\cite[p.119]{p:logic-sci-disc}.  For example, the statement ``there are white ravens'' is not falsifiable without examining every raven that exists, past, present, and future, everywhere in the universe.  Such statements are considered by Popper to be non-empirical.  On the other hand, a statement like ``all ravens are black'' is falsifiable; one need only find a single raven that is not black to refute the hypothesis.  In short, scientific criticism is enabled by testable hypotheses.

Glass, et. al. found that the majority of publications in computer science and software engineering presented abstract and formulative research findings\cite{grv:rsch-comp-disc,gvr:rsch-se}.  Hartmanis uses this emphasis on the abstract \textit{how} to justify computer science's distinction as a new kind of science\cite{h:turing}, while Stewart\cite{s:sci-cs} criticizes the vague and often unstated problems and hypotheses that characterize some areas of research in the discipline.  Tichy, Lukowicz, Prechelt, and Heinz are less gentle\citeyear{tlph:exp-eval-cs}:  they call the lack of experimental evaluation in computer science ``unacceptable, even alarming,'' echoing Jones\citeyear{j:sw-metr} characterization of the software industry as ``amateurish craft'' and an ``embarrassment.''   They discount explanations regarding the youth of the discipline, the difficulty of experimentation, and the fear of negative career impact, with the unstated implication that the root cause may be simple laziness or not wanting to figuratively rock the boat.  The risk of damage to the discipline is great, and because of the intimacy and pervasiveness of computing today, the risk to society as a whole is even greater.  To paraphrase Asimov's Zeroth Law of Robotics\cite{a:r-and-e}:  ``A computer scientist may not injure humanity, or, through inaction, allow humanity to come to harm.''

\subsection*{Human participants in computing research}\label{human}

When the topic of human participants in computer science research is discussed, the first area that probably comes to mind is human-computer interaction.  Evaluation of educational methods or analytic/modeling techniques for software development would likely follow close behind.  There are, however, other research areas that involve human participants in more subtle ways.  In this section, we consider some of these areas through discussions of specific cases.  We conclude with a comparison of corporate entities and human participants in research, and raise some hitherto unasked questions about potential ethical issues paralleling those concerning human subjects.

The first case we consider is that of research involving \textit{open source software} (OSS).  El-Emam points out that OSS is an attractive alternative for researchers desiring to analyze large software systems, as the source code is publicly and freely available\cite{e:eth-oss}.  He also points out that developers of OSS code did not intend for their work to be used as the subject of research, and thus informed consent must be given in order to use the code.  However, the task of getting consent from everyone involved in an OSS project is daunting:  there may be hundreds or even thousands of contributors, and a researcher cannot know in advance which particular individual's contributions are going to be the subject of analysis.  El-Emam also raises questions regarding minimization of harm and confidentiality.  Version control systems used in OSS projects tag contributions with information identifying the developer responsible for each piece of code integrated into the system.  If the research resulting in the publication of code segments, particularly in a negative manner, the source code could be traced back to the individual who wrote it, and possibly cause that person professional or personal harm.

Responding to this dilemma, Vinson and Singer note that eliminating personal identifiers from the reported data in order to maintain confidentiality reduces the need for informed consent, but does not necessarily eliminate it\cite{vs:source-eth}.  Furthermore, removing personal identifiers is far from simple.  Consider a popular open source e-mail client that is available for a wide variety of hardware platforms and operating systems:  Mozilla Thunderbird.  Even if the name of this particular program was not stated, the set of possible candidates is small.  Including a segment of source code could facilitate the identification of the program with a search of the OSS project's source code repository as well as the identity of the contributing developer.  Vinson and Singer also point out that removing identifiers inhibits replicating the research because subsequent researchers cannot be sure they are evaluating the same software.  This also reduces the compatibility of the initial results with later findings for the same reason.  They conclude that research with OSS is ``fraught with ethical issues,'' and encourage researchers to be proactive in the development of ethical guidelines for this type of research.

A second case is described by Storey, Phillips, and Maczewski, where they conducted research, using students as subjects, on web-based learning tools\cite{spm:web-lt-stu}.  Among the ethical issues they raised were the inadvertent coercion of students to participate (by giving course credit to participants), the fairness between students who dropped out and those who did not, and the involvement of the instructor and teaching assistant as experimenters in the research.  Participants also faced the stress of learning and switching between new and unfamiliar tools, as well as extraordinary time demands on the teaching staff due to questions about and defects in the systems under evaluation.  Davis provides extensive insight into the ethical issues in this case, focusing on the coercion that existed in the experiment\cite{d:vol}.  Davis notes that the students who ``volunteered'' for the study had registered for the class without knowing about the research.  Students registered for the class with the expectation of doing the usual course work, but were then faced with making an unexpected choice between participating in the research or writing a significant review of the tools used in the study.  Changing their schedules would have incurred significant financial cost, further coercing students into participating in the study.  While this choice is not necessarily wrong, he notes that it does require justification from two perspectives:  1) that the new requirement (participating in the study or writing the review) has positive value for the students, and 2) there is no less-coercive means to achieve the resulting good.  Davis concludes by questioning why the institution's ethics review committee did not identify the coercion in the experiment.

Sieber notes similarities between students and employees as subjects in research, in particular their vulnerability to psychological, social and economic harm\cite{s:prot-res-sub}.  Recruitment by coworkers or supervisors can be coercive; announcements should be made in public forums and volunteers accepted in private to minimize the peer pressure to participate.  Singer and Vinson identified other ethical issues involving research in corporate environments with employee subjects\cite{sv:eth-emp-stud-se}.  In the course of their research, managerial approval was required before subjects could be interviewed to ensure that the employees were not bothered during critical work periods, compromising the anonymity of the subjects.  Observation of employees at work in most corporate environments (i.e., cubicles) clearly identifies subjects to coworkers and supervisors.  Singer and Vinson also emphasized the need to work with large groups of subjects to minimize the chance of identifying individuals from research reports.  Finally, they noted that employers sponsoring research, and allowing their employees to participate in it, are often interested in how the subjects are performing.  Not only are confidentiality issues at work, but there is potential conflict of interest between the good of the research, the good of the employee/subject, and the good of the company.

Seaman raises new questions about a researcher's interactions with employees in her case regarding a pair of qualitative studies conducted at two different companies\cite{s:eth-qual-comm-sw}.  To help foster a relaxed  relationship with the subjects, the researcher (Seaman) engaged in frequent informal conversation with the employees at both companies.  The second company knew of the first study, and in the course of these informal conversations, asked questions about the work practices at the first company.  Evading the questions might reinforce the observer-subject relationship and compromise the research, while answering them might cause employees to be concerned about the researcher discussing \textit{their} work habits with a subsequent company, also risking the quality of the research.  Gotterbarn responds to this dilemma by asserting that this is not a research ethics problem, but an awkward situation, pointing out that the ethical principles are clear.  Only information specifically authorized in the consent agreement can be discussed.  Handled tactfully, the researcher can use this situation to build additional trust between herself and the company\cite{g:eth-qual-comm-sw}.

\section{Is Research Ethics a Neglected Topic in the Training of Computer Science Researchers?}\label{negl}

The purpose of this essay has been to point out some ethical issues that can occur in the course of research in computer science.  We have purposely avoided topics such as plagiarism, authorship, and more general principles of responsible research conduct, as these are common to all research areas.  Issues pertaining to professional ethics have also been omitted, as they relate to the conduct of practitioners in industry rather than researchers, topics that are better addressed in a review directed at professional practice.  It is hoped that by focusing attention on these less than obvious issues researchers in computer science and related fields will become more aware of ethical concerns in their research and begin dialogs among their colleagues on the subject.

To highlight the problem, we conclude by briefly discussing a survey conducted to investigate the awareness of research ethics among academic computer science and software engineering researchers in the United Kingdom, with discouraging results.  Among the results Hall and Flynn reported, only 16 of the 44 respondents consider monitoring the ethical considerations of software engineering research to be very important\cite{hf:eth-se-res}.  The other respondents stated they did not have feelings either way (39\%), they consider such monitoring not important (18\%), or they don't know (7\%).  Hall and Flynn also reported some striking comments from respondents:

\begin{quote}
 \begin{itshape}
 ``I find this questionnaire very worrying because the idea of having to seek ethical approval threatens academic freedom.''

 ``(Seeking ethical approval) has never arisen and I don't know why this is an issue.''

 ``No one is responsible for the ethical approval of CS research.''
 \end{itshape}
\end{quote}

This survey was conducted in the UK and the results may not be directly extrapolated to universities in the United States.  However, this author's conversations with graduate students and faculty confirms a more general apathy towards research ethics.  Many students, in particular, are not aware of review requirements for human subjects in research.  Even more discouraging, however, was the attitude towards research ethics shown by some of the candidates the author had the opportunity to question during a recent faculty search.  Several of these candidates (all were currently full professors) did not see the relevance or importance of research ethics training for PhD students in computer science, even when some of the situations discussed in this essay were described to them.  Their ethical concerns related to plagiarism and cheating.  While these are certainly important topics, there are other very significant ethical issues in computer science research as this essay has tried to illustrate.

Faulty research in computer science may not have the obvious effects that failures in medicine or other disciplines may exhibit.  Rather, these effects may be subtle and very difficult to identify when they find their way into real-world applications, but can affect the lives of millions of people.  Computers are routinely used to model and forecast weather, crop production, the spread of infectious diseases, economic markets, and many other complex systems.  Modern society relies on these models, and subtle flaws or undocumented behaviors resulting from shortcuts in research can have both immediate and long-term effects.  Algorithms can be incorporated into commonly used software tools (e.g., spreadsheets, database systems, internet search engines, etc.) based upon the performance reported in research papers, but if the real-world performance does not meet the expectations from the research, time and money are wasted.

It is not hard to imagine scenarios where computers can have significant consequences for people in the remotest parts of the world, people who have never seen a computer themselves.  Consider the following example.  A deadly disease begins spreading through small, remote villages.  Health officials use computers to store and analyze data collected on the disease, tracking its spread to identify infection vectors and characterizing its symptoms to narrow the possible causes.  If the software does not perform as expected, based on the research leading to its implementation, critical time is wasted pursuing false leads.  Other researchers use computer-aided microscopy and other analysis methods to visualize the virus and map its structure for comparison to known viruses, with more potential for lost time.    Drug companies use the collected data to guide data mining software through their repositories of compounds they have developed, searching for potential treatments, cures, and vaccines.  A data mining algorithm that was reported to work well on this kind and scale of information, based on partial testing and simulations, does not perform to expectations, but this flaw is unnoticed because the size of the data sources and complexity of the search criteria prevent manual correlation as a backup, and a probable cure goes uninvestigated.  These are not ``defects'' in the usual sense of the term --- the software does not crash, visibly erroneous results are not generated --- everything seems to be working as designed.

The pervasiveness of computers makes this possible.  Computer science is a unique discipline, and needs a unique and comprehensive approach to research ethics.  Our modern society relies heavily on computer systems that have the potential to reach into the remotest parts of the world.  Researchers must be aware of the immediate consequences of their decisions as well as the long-term and far-reaching effects they can have on the world.  Society expects this level of diligence from researchers in medicine, biotechnology, agriculture, energy, and many other disciplines.  Because computers are a vital part of virtually every other research area, computer scientists should hold themselves to these high standards and expectations.

\section{Case Study}\label{case}

Ken and Ann are doctoral students working under Dr. Smith, an internationally respected computer scientist whose specialty is the static analysis of large software systems.  Static analysis is used to analyze how a software system will behave without actually executing the program.  The results of this analysis can be used to identify programming defects and verify behavior against specifications.  Over the past decade, Dr. Smith and his students, along with colleagues from other institutions, have developed and refined a suite of tools for identifying software defects using static analysis.  Ken and Ann have each made separate modifications to detect two kinds of defects that are currently not identifiable using this suite of tools.  Their experimental setup includes ``control'' software systems written specifically to include and exclude the particular defects they are interested in, as well as mission-critical systems provided by four vendors who are also providing funding for the research in exchange for access to the resulting data.

Testing their modified tools against the control systems generates the expected results --- the intentionally introduced defects are identified, and there are no false positives.  Satisfied with these results, Ken and Ann independently run their tools against seven real-world systems, where both notice an interesting anomaly.  Their tools do indeed identify the defects, but both also return a statistically significant number of false positives from one vendor's supplied systems.

Ken attempts to duplicate the false positives by modifying the defect-free control to include the code causing the errors in the vendor systems.  All of his trials against these modified controls fail to replicate the false positives generated by the actual systems.  Stymied, Ken concludes that there must be some subtleties in other parts of the vendor's design or implementation that is causing the false positives.

Meanwhile, Ann takes a different track to identify the cause of the anomalous behavior by first running the unmodified analysis tools against the vendor systems.  Interestingly, she finds that other analyses also produce false positives with this vendor's software more consistently than with systems from the other six vendors.  Reviewing the results of earlier work with the tool suite by other researchers, Ann finds that this behavior has been noted before, but explained as an artifact of this vendor's particular programming style.  Ann contacts Ken (Dr. Smith is currently out of the country at a conference) and finds that he has noted the same behavior and concluded that the source of the problem is in the software system, not in the analysis tools.  Ann is uncomfortable with this position and suspect some kind of bias in the test suite, but a manual analysis of these software systems would be very tedious and time-consuming.  As both students are under pressing submission deadlines, they decide to write up their results as is, accepting the previous conclusions for the anomalous behavior and using the published statements as validation for their conclusions.

\newpage

\textbf{Questions for discussion:}

\begin{enumerate}
 \item The \textit{ACM Code of Ethics and Professional Conduct}\citeyear{acm:ethics} states in section 2.5 that a ACM computing professional will ``Give comprehensive and thorough evaluations of computer systems and their impacts, including analysis of possible risks.''

Similarly, the joint ACM/IEEE \textit{Software Engineering Code of Ethics and Professional Practice}\citeyear{acm-ieee:se-ethics} states that a software engineer shall:

  \begin{quote}
   \begin{itshape}
    6.07. Be accurate in stating the characteristics of software on which they work, avoiding not only false claims but also claims that might reasonably be supposed to be speculative, vacuous, deceptive, misleading, or doubtful.

    6.08. Take responsibility for detecting, correcting, and reporting errors in software and associated documents on which they work.
   \end{itshape}
  \end{quote}

In light of these statements, is it clear that Ann and Ken cannot ignore the anomalous behavior detected in the course of their research?

 \item Should anomalous results always be published?

 \item If the similar anomalies have been noted by other researchers using the same software tools, data sources, etc., should these prior comments be invoked as the sole explanation for the extraordinary behavior, or does a researcher have an obligation to extend his or her investigation in an attempt to explain the behavior?

 \item What other steps could Ken and Ann take to identify the suspected bias in their experiments?

 \item What risks do Ann and Ken expose themselves to by attempting to identify any such bias?  What are the risks if they do not?

 \item What risks exist for their advisor, Dr. Smith, if they take either course of action?

 \item What rules of thumb do we have with respect to making these decisions?

\end{enumerate}

\newpage

\section{Codes of Conduct}\label{codes}

\begin{itemize}
 \item ~ \textbf{ACM Code of Ethics and Professional Conduct}

 \hfill \url{http://www.acm.org/constitution/code.html}

 \item ~ \textbf{Software Engineering Code of Ethics and Professional Practice}

 \hfill \url{http://www.acm.org/serving/se/code.htm}

 \item ~ \textbf{IEEE Code of Ethics}

 \hfill \url{http://www.ieee.org/web/membership/ethics/code_ethics.html}

 \item ~ \textbf{AITP Code of Ethics}

 \hfill \url{http://www.aitp.org/organization/about/ethics/ethics.jsp}

 \item ~ \textbf{Australian Computer Society Code of Ethics}

 \hfill \url{http://www.acs.org.au/national/pospaper/acs131.htm}

 \item ~ \textbf{Computer Society of India Code of Ethics}

 \hfill \url{http://courses.cs.vt.edu/~cs3604/lib/WorldCodes/India.Code.html}

 \item ~ \textbf{DPMA Code of Ethics}

 \hfill \url{http://courses.cs.vt.edu/~cs3604/lib/WorldCodes/DPMA.html}

 \item ~ \textbf{An Engineer's Hippocratic Oath}

 \hfill \url{http://courses.cs.vt.edu/~cs3604/lib/WorldCodes/Hippocr.Oath.html}

 \item ~ \textbf{New Zealand Computer Society Code of Ethics \& Professional Conduct}

 \hfill \url{http://www.nzcs.org.nz/SITE_Default/about_NZCS/Code_of_ethics.asp}

 \item ~ \textbf{NSPE Code of Ethics for Engineers}

 \hfill \url{http://www.nspe.org/ethics/eh1-code.asp}

 \item ~ \textbf{American Mathematical Society Ethical Guidelines}

 \hfill \url{http://www.ams.org/secretary/ethics.html}

 \item ~ \textbf{Ethical Principles of Psychologists and Code of Conduct}

 \hfill \url{http://www.apa.org/ethics/code2002.html}

\end{itemize}

\newpage

\bibliographystyle{acmtrans}

\end{document}